\documentclass[preprint,proceedings]{rmaa}

\suppressfulladdresses % by popular request

\usepackage{paralist}

\usepackage{psfrag,color}

\SetYear{2007}
\SetConfTitle{The nuclear region, host galaxy and environment of AGN}

\title{Narrow-line Seyfert 1 Galaxies} 

\author{
  S. Komossa,\altaffilmark{1} 
  }

\altaffiltext{1}{Max-Planck-Institut fuer extraterrestrische Physik, 
       Giessenbachstrasse 1, 85748 Garching, Germany 
   (skomossa@mpe.mpg.de).}

\shortauthor{S. Komossa}
\shorttitle{NLS1 galaxies}

\listofauthors{Komossa, S.}
\indexauthor{Komossa, S.}

\abstract{I provide a short review of the
properties of Narrow-line Seyfert 1 (NLS1) galaxies across the electromagnetic
spectrum and of the models to explain them. 
Their continuum and emission-line properties
manifest one extreme form of Seyfert activity. 
As such, NLS1 galaxies may hold important clues
to the key parameters that drive nuclear activity. 
Their high accretion rates close to the Eddington rate 
provide new insight into accretion
physics, their low black hole masses
and perhaps young ages allow us to
address issues of black hole growth,
their strong optical FeII emission
places strong constraints
on FeII and perhaps metal formation models and physical conditions
in these emission-line clouds,
and their enhanced 
radio quiteness permits 
a fresh look at causes of radio loudness
and the radio-loud radio-quiet bimodality in AGN. 
   }

\resumen{Favor de proporcionar un resumen en espa\~nol. 
\\
}

\addkeyword{Galaxies: Active}
\addkeyword{Galaxies: Seyfert}

\begin{document}
% Typeset article header
\maketitle

\section{Introduction}
\label{sec:intro}

Narrow-line Seyfert 1 galaxies
are a subclass of active galactic nuclei (AGN). Their spectra  exhibit
exceptional emission-line and continuum properties.  
The most common NLS1 defining criterion is the width of the
broad component of their optical Balmer emission lines
in combination with 
the relative weakness of the [OIII]$\lambda$5007 emission 
(FWHM$_{\rm H\beta} < 2000$ km/s and [OIII]/H$\beta_{\rm totl}$  $<$ 3; 
 Osterbrock \& Pogge 1985, Goodrich 1989){\footnote{
While it is clear that a strict cutoff in line width (FWHM$_{\rm H\beta} < 2000$ km/s)
is a gross simplification of any classification scheme, this historical
value is still most commonly adopted for practical
purposes. Suggestions have been
made that more advanced NLS1 classification schemes would,
for instance, incorporate the source luminosity
(e.g., Laor 2000, Veron-Cetty et al. 2001). 
According to Sulentic et al. (2008 and references therein), AGN properties appear
to change more significantly at a broad line width of
FWHM$_{\rm H\beta} \approx 4000$ km/s.}}.  
NLS1 galaxies typically show strong FeII emission which anticorrelates
in strength with the [OIII] emission, and with the width of
the broad Balmer lines. Often the presence of  FeII emission is 
added as further NLS1 classification criterion
and Veron et al. (2001) suggest the use of an 
intensity ratio FeII/H$\beta_{\rm totl} > 0.5$. 

NLS1 galaxies as AGN with the smallest Balmer lines from 
the Broad Line Region (BLR) and the strongest 
FeII emission, cluster at one extreme end of   
AGN correlation space.
It is expected that such correlations provide some
of the strongest constraints on, and new insights in, the physical conditions
in the centers of AGN and the prime drivers of activity,
and the study of NLS1 galaxies is therefore of
particular interest. 
For instance, observations and interpretations hint at smaller 
black hole masses 
in NLS1 galaxies, and as such their black holes represent 
an important link with the elusive intermediate mass black holes, which
have been little studied so far. Accreting likely
at very close to the maximum allowed values, NLS1 galaxies are
important test-beds of accretion models. 

This paper provides a short overview of the multi-wavelength
properties of NLS1 galaxies and major models to explain them. 

\section{Emission-line and Continuum Properties}

\subsection{Multi-wavelength continuum and emission lines: trends
and correlations}

Correlations among AGN properties provide insight into the
underlying drivers and therefore enrich our understanding
of the physics and evolution of AGN.  
A commonly applied method aimed at 
identifying the strongest correlations and the underlying parameters,
is the Principle Component Analysis (e.g., Boroson 2004).
Applied to NLS1 and BLS1 galaxies,  
the strongest correlations involve the width of
the H$\beta$ line and the strength of the [OIII] line and of the FeII complex   
(e.g. Boroson \& Green 1992, Sulentic et al. 2000, 2002, Boroson 2002;
review by Sulentic et al. 2008). NLS1 galaxies show, on average,
the smallest Balmer lines, strongest FeII emission,
and smallest ratios of [OIII]/H$\beta_{\rm totl}$.  

Additional emission line trends across the NLS1-BLS1 galaxy 
population 
include 
stronger [OIII] line asymmetries of NLS1 galaxies 
(e.g., Zamanov et al. 2002, Bian et al. 2005, Boroson 2005),
stronger CIV blueshifts/asymmetries (e.g., Sulentic et al. 2000, 2007,
Wills et al. 2000, Leighly \& Moore 2004),
smaller ratios of CIV/Ly$\alpha$ (e.g., Wills et al. 2000, Kuraszkiewicz et al. 2000),
and, on average, higher intensity ratios of [SII]6716/6731
corresponding to lower densities of their
narrow-line regions (NLRs; Xu et al. 2007){\footnote{See Marziani 
et al. (2006), Sulentic et al. (2008) for reviews,
and Zhou et al. (2006) for correlation analysis and multi-wavelength
trends of the largest NLS1 sample to date, selected from the SDSS.
It has to be noted that several of the multi-wavelength trends 
mentioned in this Section are based on relatively small samples,
sometimes as small as $\sim$5 objects. }}.
In the NIR, NLS1 galaxies show higher line ratios
between coronal lines and low-ionization forbidden lines
than BLS1 galaxies (Rodriguez-Ardila et al. 2002).  

In comparison with BLAGN, NLS1 galaxies are  less often
very radio-loud (Komossa et al. 2006), tend to be overluminous in the
infrared (Moran et al. 1996, Ryan et al. 2007),  
underluminous/redder in the ultraviolett (Rodriguez-Pascual et al. 1997, Constantin \& Shields 2003),
and show a much larger scatter in X-ray spectral
slopes with, on average, steeper soft and slightly steeper hard X-ray spectra
(e.g., Puchnarewicz et al. 1992, Wang et al. 1996, Grupe 1996, Boller et al. 1996, 
Brandt et al. 1997, Leighly 1999, 
Comastri 2000, Vaughan et al. 2001, Grupe 2004, Zhou et al. 2006; 
see Williams et al. 2004 for a subsample of NLS1 galaxies with
relatively flat X-ray spectra).
NLS1 galaxies vary in a similar way as BLS1 galaxies
in the optical regime (Klimek et al. 2006), perhaps
even less than BLS1s (Ai et al. 2008, Giannuzzo et al. 1998).
Long-term optical monitoring (several decades) of a few NLS1 galaxies 
revealed variability by 0.5-1\,mag on time scales as short
as several days (Greiner et al. 1996). 
A similar coverage
is needed for much larger samples. 
On the other hand, NLS1 galaxies vary more than BLS1s in the X-ray
regime (e.g., Leighly 1999, Grupe 2004, McHardy et al. 2006).

Apart from thorough work on optical
emission-line spectroscopy{\footnote{Among NLS1 galaxies,
the prototype 
I\,Zw\,1 may have the best studied optical spectrum - see the detailed
work by Veron-Cetty et al. (2004).}}, 
NLS1 galaxies have been 
studied in greatest detail in X-rays. They show, on average, steeper
X-ray spectra than BLS1 galaxies, even though the scatter is very
large. Their X-ray spectra are complex and evidence for
cold and ionized absorption, partial covering, and reflection components has
been presented (e.g., Wang et al.
1996, Komossa \& Meerschweinchen 2000, Crummy et al. 2006,
Chevalier et al. 2006,  
Gallo 2006, Done et al. 2007, Grupe et al. 2007c).  
Which emission/absorption mechanisms dominate on average
is still being investigated.

\subsection{Frequency of NLS1 galaxies}

The observed fraction of NLS1 galaxies among AGN is a function
of wavelength band and luminosity; large samples free
of selection biases are still needed.
Soft X-ray selection turned out to be quite efficient 
in identifying NLS1 galaxies,
and Grupe (2004) reported a fraction of 46\%
NLS1 galaxies among his bright, soft X-ray selected 
broad-line AGN sample. On the other hand, the
fraction of NLS1 galaxies in other ROSAT X-ray samples was
significantly lower, and Hasinger et al. (2000) 
found only 1 NLS1 galaxy out of 69 AGN identified
in a ROSAT deep field. 
The fraction of NLS1 (vs. BLS1) galaxies in predominantly optically
selected samples is typically $\sim$15\%, but may increase in dependence
of luminosity (up to $\sim$20\%; Zhou et al. 2006).

\subsection{NLS1 models.}  

A number of scenarios have been considered in order to explain 
the observed correlations and trends in multi-wavelength
NLS1 parameter space. These include: 
Accretion rates close to or even super-Eddington
and low black hole masses (see Sect. 3 \& 4), 
winds and density effects (e.g., Lawrence et al. 1997, Gaskell 2000, Wills et al. 2000,
Bachev et al. 2004, Xu et al. 2003, 2007), metallicity (e.g., Mathur 2000, 
Shemmer \& Netzer 2002, Nagao et al. 2002, Warner et al. 2004, Shemmer et al. 2004, Fields et al. 2005), 
and
ionized absorption (e.g., Komossa \& Meerschweinchen 2000,
Gierlinski \& Done 2004, Done et al. 2007).  
While there is no space here
to review all of these in detail, I will summarize briefly
some of the major models and potential key parameters.  

\section{Black Hole Masses}

Several lines of evidence hint at small black hole masses in
NLS1 galaxies as a class.
There are different ways of determining black hole  masses of AGN in
general, and of NLS1 galaxies in particular.  
The method most commonly applied to BLAGN is 
the relation between BLR radius and black hole mass, 
based on reverberation
mapping results for Seyfert galaxies
(e.g., Peterson et al. 2004, Kaspi et al. 2005).
Few NLS1 galaxies have been reverberation-mapped so far
(Peterson et al. 2000), so an assumption often
made is that the relation obtained for BLS1  galaxies
also applies to NLS1s.
Most studies applying the mass-radius
relation, and variants of it, find systematically
lower black hole masses in NLS1 galaxies than in BLS1s 
(e.g., Boroson 2002,  
Grupe \& Mathur 2004, Collin \& Kawaguchi 2004,
Mathur \& Grupe 2005a,b,
Komossa \& Xu 2007).   
Attempts to correct for possible inclination effects
(Sect. 6) 
and/or measurements of the `second moment' of the H$\beta$ line, point
to possibly higher black hole masses (Collin et al. 2006, 
Watson et al. 2007), but still off-set from the
bulk of the BLS1 population.   
Independent evidence for
small black hole masses of NLS1 galaxies comes from X-ray observations,
particularly variability studies (e.g., Hayashida 2000,  
McHardy et al. 2006, Zhou et al. 2007b), and is based on
an observed anti-correlation between variability
time scale and black hole mass (e.g., Papadakis 2004).  
In a few cases, low black hole masses have been measured from
stellar velocity dispersion (Barth et al. 2005)
and bulge luminosity (Botte et al. 2004). 

\section{Accretion Rates}
Closely linked to (the correctness of estimates of)
black hole masses is the accretion rate relative to
the Eddington rate, generally parameterized as ratio
of $L_{\rm bol}/L_{\rm Edd}$, where the bolometric
luminosity $L_{\rm bol}$ is estimated
in most cases from measurements in a single energy band plus a  fixed
bolometric correction, and $L_{\rm Edd}=1.3\,10^{38} M/M_{\odot}$. 
Early suggestions that NLS1 galaxies might accrete close
to the Eddington rate (Boroson \& Green 1992), were 
bolstered by more recent studies, theoretical considerations,
and 
by 
estimates of 
black hole masses from optical emission-line
and continuum measurements   
(e.g., Wang et al. 1996,
Boller et al. 1996, Laor et al. 1997, Mineshige et al. 2000,
Sulentic et al. 2000,  Nicastro 2000, 
Collin \& Hur{\'e} 2001, Marziani et al. 2001, Boroson 2002, 
Kawaguchi 2003, Xu et al. 2003, Czerny et al. 2003, Collin \& Kawaguchi 2004, 
Grupe 2004, Bachev et al. 2004,
Warner et al. 2004, Tanaka et al. 2005, Collin et al. 2006, 
Xu et al. 2007).  
Independently, the shape and luminosity 
of the soft and hard X-ray spectrum was used to argue 
for accretion close to the Eddington rate,
or even super-Eddington
(e.g., Pounds et al. 1995, Kuraszkiewicz et al. 2000, 
Wang \& Netzer 2003){\footnote{Note that Williams et al. (2004) report 
a correlation between $L_{\rm 1keV}/L_{\rm Edd}$
and X-ray powerlaw index in the sense that NLS1 galaxies
with flat X-ray spectra have lower
$L_{\rm 1 keV}/L_{\rm Edd}$.}}.   
In the Eigenvector (EV) analysis of Boroson (2002), $L_{\rm bol}/L_{\rm Edd}$
drives EV1, while accretion rate drives EV2.  

\section{Winds and Outflows} 
Accretion close to the Eddington rate
implies the presence of strong radiation-pressure driven
outflows in NLS1 galaxies. Circumstancial evidence
for outflows in NLS1 galaxies on the scales of
their emission line regions comes from blue asymmetries
and shifts of spectral lines (e.g., Sulentic 2000, 2007,
Xu et al. 2003, Leighly \& Moore 2004,
Boroson 2005, Baskin \& Laor 2005, Bian et al. 2005, Aoki et al. 2005). 
Lawrence et al. (1997) speculated that the density of an outflowing
wind could be among the primary drivers of NLS1 properties.  
Indeed, Xu et al. (2007) find that the average density of the NLR
of NLS1 galaxies is lower than that
of BLS1 galaxies, and that NLR density is correlated with 
the blueshift of the blue wing of the [OIII] emission line.

\section{Inclination}

The question whether NLS1 galaxies are 
seen at preferred viewing angles -- closer 
to pole-on -- and whether this
effect could contribute to, or dominate, 
their extreme properties has been discussed 
ever since their discovery (e.g., Osterbrock \& Pogge 1985,
Puchnarewicz et al. 1992,   
Collin \& Kawaguchi 2004). 
Several lines of arguments have been presented that orientation
is not the dominant effect, for instance because
[OIII] luminosity was involved in key correlations (Boroson \& Green 1992) and
is believed to be an isotropic property.
Smith et al. (2002, 2004) have shown that polarization
properties of NLS1 and BLS1 galaxies are similar and 
argued against a pole-on orientation of NLS1 galaxies. 
On the other hand, there are some lines of evidence
that orientation might play a secondary role in 
explaining AGN emission-line correlations
and NLS1 properties in particular 
(e.g.,  Sulentic et al. 2000, Marziani et al. 2001, Boroson 2002, Bian \& Zhao 2004b, 
Zhang \& Wang 2006, Collin et al. 2006).

\section{Locus on the M-sigma plane}

Intimately linked to their high Eddington accretion 
rates and low black hole masses is the question whether NLS1 galaxies follow
the $M_{\rm BH}-\sigma$ relation of BLS1 and normal galaxies (Mathur et al. 2001).
NLS1 galaxies may hold important clues to the formation
of this relation.

Employing different methods of measuring velocity dispersion
$\sigma$ (most often by using the width of [OIII]$\lambda$5007
as substitute for stellar velocity dispersion), several authors
found that NLS1 galaxies lie off the $M_{\rm BH}-\sigma$ relation
(Mathur et al.
2001, Bian et al. 2004a, Grupe \& Mathur 2004, Mathur \& Grupe
2005a,b, Zhou et al. 2006), while other studies put them on 
the relation (Wang \& Lu 2001, Botte et al. 2005).   
Different measurements of bulge luminosity $L$ also place NLS1 galaxies
on (Botte et al. 2004) or off (Wandel 2002, Ryan et al. 2007) 
the $M_{\rm BH}-L$  relation (see Sect. 1 of Komossa \& Xu 2007 for
more detailed referencing).  
A main source of uncertainty, when examining the location
of NLS1 galaxies on the $M_{\rm BH}-\sigma$ plane, 
regards the necessary corrections which need to be applied to both,
measurements of [OIII] width and of stellar absorption line width. 
[OIII] line profiles are known to be complex
and asymmetric.  
Stellar absorption lines could be affected by  
a systematic contribution
of a disk component (and inclination effects) in non-ellipticals,
gas absorption lines by outflows.  
The width of [OIII] is most commonly used as surrogate for $\sigma$
because the bulge velocity
dispersion $\sigma_*$ is very difficult to measure
in NLS1 galaxies due to superposed emission-line complexes.
Re-examining the usefulness of the width of [OIII] as substitute for $\sigma$,
and exploring the usefulness of [SII]$\lambda$6716,6731, Komossa \& Xu (2007) found that NLS1
and BLS1 galaxies do follow the same $M_{\rm BH}-\sigma$ relation
when using the width of [SII] to determine 
$\sigma$. Furthermore, the width of the {\em core} of 
the [OIII] line is still a good surrogate for  
$\sigma$, but only after excluding objects which have their [OIII]
velocity field dominated by radial motions (presumably
outflows) as manifested in significant blueshifts of these [OIII]
core lines (Fig. 1).  

\begin{figure}[!t]
  \includegraphics[width=\columnwidth]{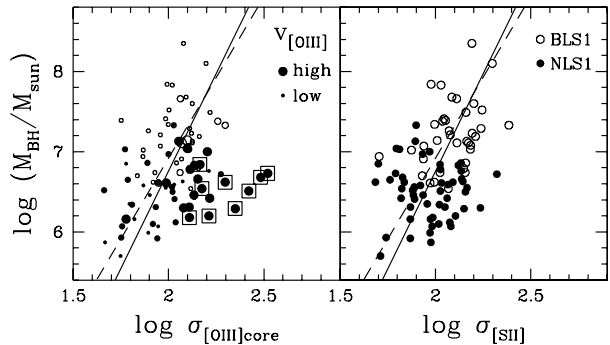}
  \caption{Location of NLS1 galaxies (filled symbols) and BLS1
galaxies (open symbols) on the $M_{\rm BH}-\sigma$ plane (Komossa \& Xu 2007). 
The {\bf{left panel}} is based on $\sigma$ measurements from the narrow core of [OIII]$\lambda$5007
(asymmetric blue wings were removed). Circle size is coded according to
the blueshift of the core of [OIII]. `Blue outliers' in [OIII] (with
radial velocities larger than 150 km/s) are marked with an extra open square.
Once these blue outliers are removed, NLS1 and BLS1 galaxies follow the same
$M_{\rm BH}-\sigma_{\rm{[OIII]}}$ relation. The {\bf{right panel}} shows the same relation based on [SII]
(without any velocity coding; high outflow velocities
only appear in [OIII]).  NLS1 and BLS1 galaxies follow the same $M_{\rm BH}-\sigma_{\rm{[SII]}}$ relation.
The dashed and solid lines represent the $M_{\rm BH}-\sigma_*$ relation of non-active
galaxies of Tremaine et al. (2002) and
of Ferrarese \& Ford (2005),
respectively. $\sigma$ is measured in km/s.   }
  \label{fig1-komossa}
\end{figure}

\section{Host Galaxies}

Relatively little is known about the host galaxies of NLS1 galaxies
as a class. Krongold et al. (2001) studied host galaxies and environment
of a sample of 27 NLS1s and concluded that they reside in similar
environments as BLS1 galaxies, and tend to have smaller host galaxies.
There are indications that the fraction of bars is higher in
NLS1 galaxies than in  
BLS1s (Crenshaw et al. 2003, Ohta et al. 2007).
Among galaxies with large-scale bars, NLS1 galaxies show a higher fraction
of nuclear dust spirals and stellar nuclear rings which might indicate
more efficient fueling of their black holes (Deo et al. 2006). 
NIR imaging of NLS1 host galaxies (selected to be of type E or S0) 
revealed that their bulges
are redder than comparison samples of non-active and BLS1
galaxies (Ryan et al. 2007; see also Rodriguez-Ardila \& Mazzalay 2006).

\section{Radio-loudness} 

Until recently, relatively little was known
about the radio properties of NLS1 galaxies (Ulvestadt et al. 1995). 
Correlation analyses showed that radio loudness preferentially occurs
in objects with broad Balmer lines and weak FeII emission
(e.g., Boroson \& Greene 1992, Sulentic et al. 2003), leading to
expectations that radio loudness in NLS1 galaxies
might be rare.
More studies of radio-loud NLS1 galaxies were needed
to test this and to  
shed new
light on our understanding of the NLS1 phenomenon, 
but also on radio loudness in general since
a major unsolved question in the study of AGN
concerns the key parameters
that drive radio loudness (e.g.,
 Wilson \& Colbert 1995, Best et al. 2005, 
Capetti \& Balmaverde 2006, Sikora et al. 2007). 
A systematic search
for {\em radio-loud} NLS1 galaxies (Komossa et al. 2006)
has shown that NLS1 galaxies as a class are
not completely radio-quiet, but they are less
likely to be radio loud than BLAGN.
Only 7\% of all NLS1 galaxies
are radio loud (Komossa et al. 2006, Zhou et al. 2006), 
while only $\sim$2.5\% of
the NLS1s are `very' radio loud (radio index $R > 100$){\footnote{The fraction 
of radio-loud NLS1s in a purely radio-selected sample 
(Whalen et al. 2006) is higher, but even in that sample very
radio loud objects are rare.}}. 
The radio-loud NLS1 galaxies are generally
compact,  steep spectrum sources
in the radio regime and as such they share
some similarities with the previously known
class of compact steep-spectrum (CSS) radio sources.
Estimates of black hole masses show
that the radio-louds are at the upper range
of NLS1 black hole  masses, and are located in a previously scarcely
populated area of $M_{\rm BH}-R$ diagrams.
Optical properties of these radio-loud NLS1 galaxies
are similar to the NLS1 population as a whole
(Komossa et al. 2006; the same holds for a radio-selected NLS1 sample
of Whalen et al. 2006).
A surprising exception is that radio-loud NLS1 galaxies
preferentially show moderate--strong FeII emission
when compared to radio-quiet NLS1s (Komossa et al. 2006).
Mechanisms which drive the NLS1
radio properties (accretion rate, black hole spin, host
galaxy properties and merger history) are still being
explored.    

There is recent evidence that several of the radio-loudest NLS1 galaxies
display blazar characteristics and harbor relativistic jets
(Zhou et al. 2007a, Doi et al. 2007, Yuan et al. 2007).  

\section{Links with other types of sources}

\paragraph{BAL quasars.}
Similarities
between broad absorption line (BAL) quasars and NLS1 galaxies, 
mostly based on their optical line properties,
have been repeatedly pointed out   
(e.g., Brandt \& Gallagher 2000, Boroson 2002, Grupe et al. 2007b, and references therein).  
Boroson (2002) finds that the two source populations,
NLS1 galaxies and BAL quasars, are at similar ends
of his EV1, but at opposite ends of EV2 (his Fig. 7) which he
interprets in terms of different accretion rates.  
A possible Rosetta stone 
to test such links is the NLS1 galaxy WPVS007, which showed
a dramatic drop in its X-ray flux (Grupe et al. 1995, 2007a)
that appears to be accompanied by the onset of BAL activity in the UV
(Leighly et al. 2006).  

\paragraph{Galactic black hole candidates.}
 Pounds et al. (1995) pointed out the similarity of 
the X-ray spectrum of the NLS1 galaxy RE1034+39 to the high-state of  Galactic black hole
candidates, which led to more general speculations
about links between these systems.  
Similarities in spectral and
temporal properties between these two types
of systems exist in the X-ray regime (e.g., Zycki 2001, Negoro 2004, McHardy et al. 2006)
and perhaps with regard to their radio properties (Komossa et al. 2006), 
but further work is needed to explore how far these similarities go.  

\paragraph{Evolutionary
  sequences.} 
Links of NLS1 galaxies with other types of AGN have also
been suggested in the context of evolutionary sequences.  
Mathur (2000) presented lines of evidence that NLS1
galaxies might be in an early stage of evolution.
Kawakatu et al. (2007) discussed
connections between NLS1 galaxies and ultraluminous infrared galaxies
(ULIRGs; see also Hao et al. 2005) and proposed an evolutionary
sequence from type 1 ULIRG to NLS1 to BLS1.   
Zhang \& Wang (2006; see also Haas et al. 2007) suggest a link between NLS1 galaxies
and non-HBLR Seyfert\,2 galaxies (i.e., Sy 2 galaxies which lack a BLR
even in polarized light), in the sense that NLS1 galaxies are the face-on equivalents of
non-HBLR Sy2 galaxies, and Wang \& Zhang (2007) discuss the temporal
evolution of these systems into BLAGN.  

\section{Summary} 

-- {\sl{It seems hard to resist the feeling that nature is telling
us something important here, but we do not yet know what it is.}} 
~~~~~~~~~~~~~~Lawrence et al. (1997)  
\vskip0.3cm

While our knowledge has increased significantly in the last decade,
important questions are still open.
For instance, what are sufficient, what are necessary conditions
for the onset of NLS1 activity ? 
For instance, the question is raised whether there are 
two types of Seyfert galaxies with low black hole masses:
there is the unavoidable low-black-hole mass extension of BLS1 galaxies.
Such systems would have their FWHM$_{\rm H\beta}$ fall below the formal cutoff
value of 2000 km/s.  
Does low black hole mass already imply the emergence of some or all of the typical 
observed NLS1 characteristics ? Or is there
a separate class of NLS1 galaxies ?  While many
individual NLS1 galaxies have been studied in great detail,
we still need larger samples free of selection biases
and well-suited BLS1 comparison samples in order to identify
robust trends. Correlation space will ultimately have 
to be expanded to include 
the radio and infrared properties of NLS1 galaxies,
as well as the properties of their host galaxies. 
On the theoretical/modeling side, interesting questions that persist
or have emerged
are related to mechanisms of super-Eddington accretion,
the simultaneous presence of (TeV) blazar activity and
high accretion rate in extreme radio loud NLS1 galaxies,
and mechanisms of fueling and feedback in NLS1 galaxies.
The study of NLS1 galaxies will continue to provide important contributions
to our understanding of AGN and their cosmic evolution.

\acknowledgments
{\sl Acknowledgements.} It is my pleasure to thank the organizers
of the conference for their invitation and for their excellent organization
and the pleasant atmosphere, and 
the participants for many enlightening discussions.

\end{document}